\documentclass[letterpaper]{article} 
\usepackage{aaai2027}  
\usepackage[hyphens]{url}  
\usepackage{graphicx} 
\usepackage{natbib}  
\usepackage{caption} 
\newtheorem{theorem}{Theorem}
\usepackage{algorithm}
\usepackage{algorithmic}
\usepackage{amsmath,amssymb}
\usepackage{multirow}
\usepackage{graphicx}  
\usepackage{subcaption}
\usepackage{hhline}
\usepackage{booktabs}
\usepackage{multirow}
\usepackage{graphicx}
\usepackage{makecell}
\usepackage[table]{xcolor}

\usepackage{newfloat}
\usepackage{listings}
\DeclareCaptionStyle{ruled}{labelfont=normalfont,labelsep=colon,strut=off} 
\floatstyle{ruled}
\newfloat{listing}{tb}{lst}{}
\floatname{listing}{Listing}

\usepackage{booktabs}

\title{
SUCRe: Selective Uncertainty-Aware Contrastive Representation for \\ Graph Transfer Learning 
} 
\author{
    Mingcan Wang\textsuperscript{\rm 1},
    Junchang Xin\corresponding \textsuperscript{\rm 1},
    Zhongming Yao\textsuperscript{\rm 2},
    Bing Tian Dai\textsuperscript{\rm 3},
    Kaifu Long\textsuperscript{\rm 1},
    Zhiqiong Wang\textsuperscript{\rm 4}. 
}
\affiliations{
    \textsuperscript{\rm 1}School of Computer Science and Engineering, Northeastern University.\\
    \textsuperscript{\rm 2}The Department of Computer Science, Aalborg University.\\
    \textsuperscript{\rm 3}School of Computing and Information Systems, Singapore Management University.\\
    \textsuperscript{\rm 4}College of Medicine and Biological Information Engineering, Northeastern University.\\

}

\begin{document}
\maketitle

\begin{abstract}
Graph transfer learning (GTL) provides a promising paradigm for adapting knowledge from source graphs with sufficient labels to label-scarce target graphs. However, existing approaches often assume that transferred knowledge is uniformly reliable, ignoring the different transferability of samples caused by structural and distribution shifts across graphs. This limitation leads to negative transfer and unnecessary computational overhead.

In this work, we propose SUCRe, a selective uncertainty-aware contrastive representation method for GTL. The key idea is to selectively adapt and transfer graph knowledge according to its estimated reliability. Specifically, we introduce structure-aware entropy-based matching discrepancy, which jointly models feature uncertainty and structural coherence to ensure accurate feature adaptation between graphs. Moreover, we develop a domain-aware semi-hard negative sampling strategy that constructs informative contrastive sets by filtering unreliable cross-domain relationships, reducing computational redundancy while enhancing representation discrimination. Extensive experiments on graph transfer benchmarks demonstrate that SUCRe achieves competitive performance with improved efficiency. 
\end{abstract}


\section{Introduction}
Graph Neural Networks (GNNs) have been demonstrated as a powerful and versatile paradigm for modeling and analyzing complex graph-structured data \cite{GNN03, GNN05, GNN07, TCBB_my}. By leveraging message-passing mechanisms to aggregate information from neighboring nodes, GNNs achieve remarkable performance across node-, edge- and graph-level tasks in various domains (e.g. social network \cite{social01, social03, dasfaa_my}, traffic network \cite{traffic01, traffic03} and bioinformatics \cite{bio01, bio03, bio05}). Nevertheless, realizing full potential of GNNs in practice remains challenging, among which the issue of label sparsity is a fundamental bottleneck limiting the deployment of GNNs in real-world scenarios. In the cases, supervised GNNs often exhibit severely degraded performance, as they often suffer from insufficient representation learning when labels are sparse \cite{sparse, sparse4}. 

An effective solution to this is GTL, whose general paradigm is to learn transferable knowledge from a source graph and adapt it to a target graph, by leveraging shared structural patterns and feature representations to improve learning performance under limited labels on target graph. Common techniques in graph transfer learning include multi-task learning \cite{mt01, mt03}, multi-network learning \cite{uniGNN, GCC}, domain adaptation \cite{CDNE, da03} and pre-train fine-tune approaches \cite{GraphControl, ptft03}. 
In spite of their conceptual appeal, effective and efficient GTL is non-trivial. Distribution shifts between source and target graphs—manifested in both node features and underlying topological structures—often lead to negative transfer. Thus, the transferred model may even under-perform training from scratch on the target graph. 

To tackle the challenges, increasing efforts have been witnessed recently. For instance, Qiu {\it et al.} \cite{GCC} proposed a self-supervised GNN pre-training framework to capture the universal network topological properties across multiple networks. Lin {\it et al.} \cite{uniGNN} proposed a unified GNN pre-training method for multi-domain graphs to overcome the `one-domain-one-model' limitation. While these methods tend to learn a unified pre-trained model from many graphs, another branch of methods focuses on source-specific pre-training for graph-to-graph transferring, which can preserve fine-grained structural knowledge and avoid over-generalized representations introduced by multi-graph pre-training. For example, Yang {\it et al.} \cite{GraphLoRA} proposed an effective parameter-efficient method for transferring GNN pre-trained on a single source graph to target graph in a graph-to-graph manner. 


Despite the preceding successes, how to effectively adapt well-trained GNNs to graphs with sparse labels remains under-explored, particularly due to the following challenges.

\textbf{\textit{Challenge I: Cross-graph distribution mismatch due to uncertainty.}} Despite the remarkable progress in GTL, existing methods still suffer from limited transferability and limited utilization of information due to the inherent distribution discrepancy between source and target graphs. Concretely, existing works \cite{GTOT, GraphControl} always treat all transferred samples equally, ignoring the uncertainty caused by graph difference. Consequently, achieving robust generalization, especially under few-shot target settings, remains challenging.

\textbf{\textit{Challenge II: Extensive transferring computational burden.}} 
While one of the foremost aims of GTL is to train GNNs in the presence of sparse labels, computational cost is an issue that cannot be bypassed. Current GTL methods \cite{GraphLoRA, AdapterGNN} always suffer from extensive computational burden, including both time consumption and GPU memory load. Meanwhile, complex designs for ungrading GTL accuracy usually result in larger memory load and higher time cost. Conversely, reducing computational overhead via model simplification or others tends to sacrifice model accuracy.

To tackle the challenges, SUCRe is proposed in this paper, which is a \textbf{S}elective \textbf{U}ncertainty-aware \textbf{C}ontrastive \textbf{Re}presentation method for GTL. Inspired by GraphLoRA \cite{GraphLoRA}, SUCRe also construct a small trainable GNN alongside the pre-trained frozen one, while making some fundamental innovations on feature adaptation and contrastive learning. To be specific, an entropy-based feature adaptation is designed, which can achieve effective and efficient node feature adaptation from source to target graphs considering uncertainty. Furthermore, semi-hard negative sample selection is introduced into SUCRe to improve training effectiveness and efficiency, for neglecting the samples that are too hard or too easy. As a result, SUCRe achieves better overall accuracy and efficiency upon various competitive baselines. Overall, the contributions of ours can be summarized as: 

\begin{itemize}
    \item We propose Structure-aware Entropy-based Matching Discrepancy (SEMD) for GTL feature adaptation, which could align the global information between graphs considering uncertainty and preserve the structural coherence. 
    \item We propose a semi-hard negative sampling strategy that can omit the samples that are less informative or reliable, to upgrade both effectiveness and efficiency. 
    \item We conduct extensive experiments, which demonstrates the effectiveness and efficiency of the overall SUCRe framework and each proposed module. 
\end{itemize}

\section{Related Work} 
\subsection{Graph Transfer Learning} 
GTL can be mainly categorized into: multi-task learning, multi-network learning, domain adaptation, and pre-train fine-tune approaches. Compared to the others, pre-train fine-tune approaches provide a more practical solution by enabling knowledge transfer from pre-trained models to downstream tasks with minimal supervision, which consists of two steps, pre-training and fine-tuning. Graph pre-training aims to learn generalizable representations from large-scale graph data. In addition to the pre-training methods mentioned in the introduction, common pre-training methods also include PHE\cite{PHE}, GPCF \cite{GPCF} and RETRACTED \cite{RETRACTED}. Another important step of pre-train fine-tune approaches focuses on fine-tuning, aiming to adapt pre-trained GNNs to various graphs. Recent studies show that fine-tuning pretrained GNNs can improve transferability and prevent catastrophic forgetting under label scarcity and domain shift. These methods include GTOT \cite{GTOT}, AdapterGNN \cite{AdapterGNN} and GraphControl \cite{GraphControl}. 

Meanwhile, feature adaptation (a.k.a. domain adaptation) is an important concept in many classes of GTL, that utilizes the knowledge of relevant source domain(s) to assist learning task in the target domain. CDNE \cite{CDNE} is one of the pioneer works which learn transferable node embeddings for cross network learning tasks by minimizing the maximum mean discrepancy (MMD) loss. But this function neglects the structural information, which limit its adaptation performance. Recent work \cite{GraphLoRA} extended MMD to structural-aware maximum mean discrepancy (SMMD) to limit the feature discrepancy between graphs in consideration of the structure information, while neglecting the uncertainty of nodes and features. 

Nevertheless, no matter MMD/SMMD-based domain adaptation methods or existing graph fine-tuning approaches always treat the structures and features of source and target graphs in a deterministic and uniform manner, thereby largely overlooking the uncertainty in graphs. Compared to them, our work characterize transfer uncertainty by measuring the entropy of cross-domain similarity distributions to enable uncertainty-aware feature adaptation by distinguishing confident transferable samples from ambiguous or noisy ones, while preserving the structural coherence. This helps mitigate negative transfer under graph domain shift and improves the robustness of graph transfer learning. 

\subsection{Semi-hard Negative Sample Selection}

Semi-hard negative sampling has proven to be an effective technique in contrastive learning. Originally introduced in FaceNet \cite{schroff2015facenet}, the method selects negatives that are farther from the anchor than positives but still lie within a challenging margin, striking a balance between training stability and discriminative power. This idea has since been widely adopted and extended \cite{kalantidis2020hard, ma2023dropmix, xie2023negative}. These works collectively demonstrate that semi-hard negative sampling is a versatile and powerful paradigm, consistently enhancing the quality of learned embeddings by focusing on informative boundary cases while avoiding the instability of overly hard negatives. 

Despite the success of semi-hard negative sampling, its exploration in graph contrastive learning remains limited. Unlike images and texts, graphs are characterized by irregular topology and complex interactions, making the identification of semi-hard negatives even more challenging. Motivated by this, in this work, a domain-aware semi-hard negative sample detection method is proposed which filters the negative samples that are less informative or reliable. This is because that learning from less informative samples results in little impact on the gradient update of the model, thus wasting training resources and time, while excessively difficult negatives may be unreliable and introduce noisy supervision, especially under graph domain shifts where structural similarity does not always correspond to semantic similarity. 

\begin{figure*}[!t]
\centering
\includegraphics[width=0.93\textwidth]{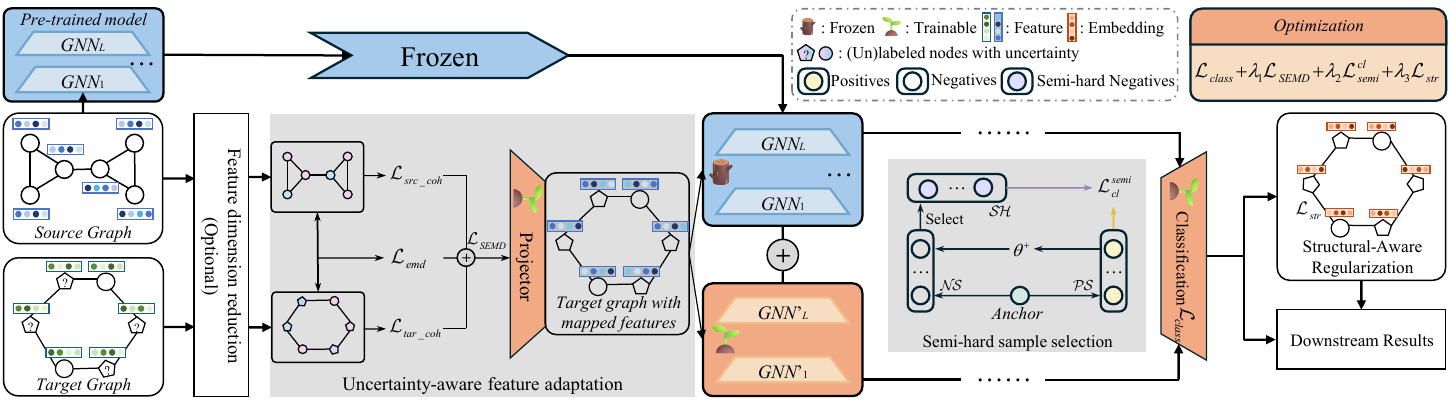} 
\caption{Selective uncertainty-aware contrastive representation method for GTL.}
\label{fw}
\end{figure*}

\section{Methodology}
\subsection{Notation and Preliminaries}

\textbf{Notation}. In this work, we utilize $G=\left(V, E\right)$ to denote a graph, where $V=\{v_1, v_2, ...,  v_N\}$ denotes the node set and $E=\{(v_i, v_j), v_i, v_j \in V\}$ presents the edge set in graph $G$. The features of $N$ nodes are denoted as $\textbf{X} \in \mathbb{R}^{f\times N}$, with its element $\textbf{x}_i \in \mathbb{R}^{f}$ representing the node features of node $v_i$. And $f$ is the dimension of features. For distinction, we use superscript `$s$' and `$t$' to present source and target graph, respectively. Specifically, $\textbf{X}^s$ means the feature matrix of the source graph, while $\textbf{x}_i^t$ means the node features of node $v_i$ of the target graph. Furthermore, denote $D_{i,i} = \sum\nolimits_{j=1}^{{{N}}}{{{A}_{i,j}}}$, where $A$ represents the adjacency matrix of the considered graph. Then the formal personalized PageRank (PPR) is formalized as:

\begin{equation}\label{pprmat}
    \boldsymbol{S} = \alpha \left(I - \left(1-\alpha\right) D^{-1/2}A D^{-1/2} \right)^{-1}
\end{equation}
where $\alpha \in \left(0,1\right)$ is the teleport probability. Each element $S_{i,j}^t$ reflects the connection between $v_i$ and $v_j$ in the target graph $G^t$.  


\textbf{Problem Definition}. Given the source graph $G^s$, target graphs $G^t$ and a GNN $g_{ptr}$ pretrained on source graph $G^s$, the objective is to train a combined GNN $\mathcal{F} = g_{ptr} \circ g_{tune} \left( \cdot \right)$, 

\begin{equation}
    \hat{\mathcal{F}} = \mathop{\arg \min } \mathcal{L}\left( \mathcal{F} \left( G^s, G^t, \textbf{X}^s, \textbf{X}^t \right), Y_{train}^t \right)
\end{equation}
where $Y_{train}^t$ is the train labels from the target graph and $\mathcal{L}$ is the model loss. 

\subsection{Framework Overview}

The overall framework of SUCRe is shown in Fig. \ref{fw}. We first perform SVM-based node feature dimension reduction on source, target or both graphs. And this step is optional. Next, we introduce the first component of SUCRe, namely uncertainty-aware feature adaptation. Specifically, we employ SEMD ($\mathcal{L}_{SEMD}=\mathcal{L}_{EMD}+\mathcal{L}_{src\_coh}+\mathcal{L}_{tar\_coh}$) to minimize the discrepancy between the feature distributions of the source and target graphs. By incorporating entropy-based uncertainty estimation, this metric not only involves uncertainty while mapping features, but also preserves the structural coherence of both graphs.

Simultaneously, a domain-aware semi-hard negative sample detection strategy is introduced to ensure selective graph contrastive learning, whose objective is to eliminate uninformative and unreliable contrastive pairs, thereby improving representation discriminability while reducing unnecessary computational overhead. During training, we keep the pre-trained GNN $g_{ptr}$ frozen and only optimize an additional GNN for efficient adaptation. Finally, we apply the structure-aware regularization to further enhance the adaptability of the pre-trained GNN for downstream tasks. 

\subsection{Structure-Aware Entropy-Based Matching Discrepancy}

Existing distribution alignment methods, such as MMD and SMMD,
mainly align feature statistics while treating node representations
in a deterministic and uniform manner. However, under cross-graph
distribution shifts, different nodes may exhibit substantially
different transfer reliability. To characterize such uncertainty,
we define a structure-aware entropy representation for each node
and align the corresponding entropy distributions between the
source and target graphs.

Let
\begin{equation}
    \mathbf{z}^{s}_{i}=\mathbf{x}^{s}_{i},
    \qquad
    \mathbf{z}^{t}_{i}=p(\mathbf{x}^{t}_{i};\boldsymbol{\omega}),
    \label{eq:feature_mapping_revised}
\end{equation}
where $p(\cdot;\boldsymbol{\omega})$ is the trainable feature
projector and $\mathbf{z}^{t}_{i}$ denotes the mapped feature of
target node $v_i^t$. For each domain $d\in\{s,t\}$, we define the local similarity
distribution of node $i$ as
\begin{equation}
    q^{d}_{ij}
    =
    \frac{
        \exp\left(
        \operatorname{sim}
        (\mathbf{z}^{d}_{i},\mathbf{z}^{d}_{j})/\tau
        \right)
    }{
        \sum_{k\in\mathcal{C}^{d}_{i}}
        \exp\left(
        \operatorname{sim}
        (\mathbf{z}^{d}_{i},\mathbf{z}^{d}_{k})/\tau
        \right)
    },
    \label{eq:local_similarity_distribution}
\end{equation}
where $\operatorname{sim}(\cdot,\cdot)$ denotes cosine similarity,
$\tau>0$ is the temperature parameter, and
$\mathcal{C}^{d}_{i}$ denotes the candidate node set used to
estimate the local similarity distribution. It can contain all nodes
or a sampled structural neighborhood. The structural-feature uncertainty of node $i$ is measured by
the entropy
\begin{equation}
    H^{d}_{i}
    =
    -\sum_{j\in\mathcal{C}^{d}_{i}}
    q^{d}_{ij}\log q^{d}_{ij}.
    \label{eq:node_entropy_revised}
\end{equation}

A small entropy value indicates that the representation of node
$i$ is concentrated around a small number of structurally or
semantically similar nodes, whereas a large entropy value indicates
an ambiguous local representation.

To incorporate global graph structure, we first compute the Personalized PageRank matrix according to Eq. \ref{pprmat}. Let $H^s = \{H_i^s \mid i \in \mathcal{V}^s\}$ and $H^t = \{H_i^t \mid i \in \mathcal{V}^t\}$ be the sets of node entropies from the source and target graphs, respectively. The cross-graph entropy-based feature adaptation loss could be easily defined as:

\begin{equation}
    \mathcal{L}_{\text{emd}} = \left\| \mathbb{E}[H^s] - \mathbb{E}[H^t] \right\|_2^2
\end{equation}
where $\mathbb{E}[H]$ denotes the average of $H$, $\left\| \cdot \right\|_2$ represents the $Euclidean$ norm.   

Subsequently, we introduce structure-aware weights based on the Personalized PageRank (PPR) matrix $\boldsymbol{S}^d = \{S_{i,i'} \}$:

\begin{equation}
    \gamma_{i,i'} = \log\left(1 + 1/{S_{i,i'}}\right)
\end{equation}
Then the final Structure-aware Entropy Mean Discrepancy loss function as follows: 

\begin{align}
\mathcal{L}_{\text{semd}} &= \sum\limits_{i,i'}{{{\gamma}_{i,i'}}{{\left( H_{i}^{t}-H_{i'}^{t} \right)}^{2}}} \\ \nonumber
& + \left\| E[{{H}^{s}}]-E[{{H}^{t}}] \right\|+\sum\limits_{i,i'}{{{\left( H_{i}^{s}-H_{i'}^{s} \right)}^{2}}} \label{eq_semd}
\end{align}
where the middle term aligns the global information between source and target graphs, whereas the other terms preserve the structural coherence within the graphs. The existence of $\gamma_{i,i'}$ (decreases as the growth of PPR) is to avoid the mapped features of the similar nodes (with higher PPR) in target graphs too `distant', thus preventing the complete collapse of the mapped features. By minimizing $\mathcal{L}_{\text{semd}}$, we can train the parameters $\boldsymbol{\omega}$ in formula (\ref{eq:feature_mapping_revised}) to project the node features in consideration of uncertainty. 

%

\begin{theorem}

Given ${G}^{s}$ and ${G}^{t}$ denote the source and target graphs, the source-target risk discrepancy satisfies: 
\begin{equation}
\left|
R^t(f)-R^s(f)
\right|
\leq
\lambda^{\star}
+
L_H C_{\mathrm{str}}
\sqrt{\mathcal{L}_{\text{semd}}},
\label{addd1}
\end{equation}
Thus, by minimizing $\mathcal{L}_{\text{semd}}$, we can limit the source-target risk discrepancy, thus ensuring effective uncertainty-aware feature adaptation. Proof and descriptions of parameters of Eq. \ref{addd1} is provided in Appendix A. 
\end{theorem} 

\subsection{Semi-Hard Negative Sample Detection for Structural Transfer. }

Inspired by previous work \cite{GraphLoRA, GraphControl}, we rethink the computational efficiency of the adaptation of pre-trained model to the target graph. Notate $g_{ptr}$ as the frozen GNN model and $g_{tune}$ as the add GNN model to be tuned on target graph. To be parameter efficient, inspired by GraphLoRA \cite{GraphLoRA}, we train the $g_{tune}$ in low-rank adaptation (LoRA) manner. Let $GNN_{ptr}^l\left(\cdot, W^{l}\right)$ and $GNN_{tune}^l\left(\cdot, W^{l*}\right)$ denote the $l$-th layer output of $g_{ptr}$ and $g_{tune}$, where $W^l$ and $W^{l*}$ are parameter matrices. For adapting, we combine the output of  $g_{ptr}$ and $g_{tune}$ at $l$-th layer as:  

\begin{align}
    \mathcal{H}^l =& GNN_{ptr}^l\left(\mathcal{H}^{l-1}, W^{l}\right) + GNN_{tune}^l\left(\mathcal{H}^{l-1}, W^{l*}\right) \\ \nonumber
    =& GNN_{ptr}^l\left(\mathcal{H}^{l-1}, W^{l}\right) + GNN_{tune}^l\left(\mathcal{H}^{l-1}, W_{A}^{l}W_{B}^l\right)
\end{align}
where $W_{A}^{l} \in \mathbb{R}^{d^{l-1}\times r}$, $W_{B}^l \in \mathbb{R}^{r \times d^{l}}$ ($r \ll min\left(d^{l-1}, d^{l}\right)$) and $\mathcal{H}^0 = \mathbf{Z}^t$ is discussed in Eq. (\ref{eq:feature_mapping_revised}). 

Then we kept the pre-trained GNN $g_{ptr}$ frozen and train $g_{tune}$ via contrastive learning. The embeddings of nodes belonging to the same
class are considered as positive samples, constructing positive set $\mathcal{PS}$, and the embeddings of nodes from different classes are treated as negatives, constructing negative set $\mathcal{NS}$. However, it is our observation that not all of the negative samples contribute to the model training in graph transfer learning. Easy negative samples provide limited discriminative information due to their clear dissimilarity from positive samples, leading to inefficient optimization. Conversely, extremely hard negative samples may introduce ambiguity and noise, causing the model to learn misleading patterns. The finding agrees with the numerous studies on semi-hard negative sample mining \cite{schroff2015facenet, kalantidis2020hard, ma2023dropmix, xie2023negative}. 



To further enhance the effectiveness and efficiency of structural knowledge transfer, we propose a domain-aware semi-hard negative sampling strategy to achieve effective and efficient graph contrastive learning. In graph contrastive learning, semi-hard negatives can encourage the model to align embeddings with label-consistent structures while preserving feature-level discriminability, thereby achieving effects analogous to supervised or downstream-task-aware training. This approach retains label supervision information while focusing on the most informative negative samples, thereby providing more stable and effective gradient signals. 


Let ${\mathbf{e}_{ptr}}=\{{e}_i\}_{i=1}^N$ and $\mathbf{e}_{tune}=\{{e}'_i\}_{i=1}^N$ denote the node embeddings produced by the frozen pre-trained GNN and the trainable LoRA-augmented GNN on the target graph, respectively. We first apply $\ell_2$-normalization to all embeddings:
\begin{equation}
\tilde{\mathbf{e}}_{ptr} = \frac{\mathbf{e}_{ptr}}{\|\mathbf{e}_{ptr}\|_2}, \quad 
\tilde{\mathbf{e}}_{tune} = \frac{\mathbf{e}_{tune}}{\|\mathbf{e}_{tune}\|_2}.
\end{equation}

For each anchor embedding ${\mathbf{e}^{anchor}}$, we define its positive sample $\mathcal{P}_{anchor}$ as:
\begin{equation}
\mathcal{P}_{anchor} = \{ j \mid label_{anchor} = label_j \},
\end{equation} 
which includes all nodes with the same label as the anchor and the anchor itself. Then the pairwise similarity is computed using temperature-scaled cosine similarity:
\begin{equation}
\theta_{ij} = \frac{\tilde{{e}}_i ^\top \tilde{e}'_j}{\epsilon},
\end{equation}
where $\epsilon$ is the temperature hyperparameter.

We dynamically select a set of semi-hard negatives $\mathcal{SH}_i$ for each anchor $i$:
\begin{equation}
\mathcal{SH}_i = \Big\{ j \notin \mathcal{P}_i \ \Big|\ 
\theta_i^+ - m < \theta_{ij} < \theta_i^+ + \delta \Big\},
\end{equation}
where $\theta_i^+ = \max_{k \in \mathcal{P}_i} \theta_{ik}$ is the maximum similarity between the anchor and its positive samples, $m > 0$ is the margin hyperparameter, and $\delta$ is a relaxation upper bound (typically set between $0.3$ and $0.5$). If $\mathcal{SH}_i$ is empty, we fall back to selecting the Top-$K$ hardest negatives.

As a result, the loss for a single anchor $i$ is defined as:

\begin{equation}
\ell_i = -\log\left( \frac{\sum_{k \in \mathcal{P}_i} \exp(\theta_{ik})}{\sum_{k \in \mathcal{P}_i} \exp(\theta_{ik}) + \sum_{j \in \mathcal{SH}_i} \exp(\theta_{ij})} \right).
\end{equation}

The final bidirectional structural contrastive loss is:
\begin{equation}
\mathcal{L}_{cl}^{\text{semi}} = \frac{1}{2N} \sum_{i=1}^N \Big( \ell_i({\mathbf{e}_{ptr}}, {\mathbf{e}_{tune}}) + \ell_i({\mathbf{e}_{tune}}, {\mathbf{e}_{ptr}}) \Big).
\end{equation}

\subsection{Optimization} Finally, a classifier $\hat{y}_i = c\left(\mathcal{H}_L\right)$ is employed for node classification, where $c\left(\cdot\right)$ represents the classification function. And the classification loss is defined as: 

\begin{equation}
    \mathcal{L}_{class}=-\frac{1}{N^t}\sum\limits_{i}{\left[y_ilog\left(\tilde{y}_i\right)+\left(1-y_i\right)log\left(1-\tilde{y}_i\right)\right]}
\end{equation}

Besides, inspired by previous work , a structure regularization term $\mathcal{L}_{str}$ based on the homophily principle of graph data is applied while optimizing. Detailed information of $\mathcal{L}_{str}$ could be referred to \cite{str, GraphLoRA}. Then the overall training objective function is defined as: 

\begin{equation}
\mathcal{L}=\mathcal{L}_{class}
+\lambda_{1}\mathcal{L}_{semd}
+\lambda_{2}\mathcal{L}_{cl}^{\text{semi}}
+\lambda_{3}\mathcal{L}_{str},
\tag{15}
\end{equation}
with each term discussed above and $\lambda_1$, $\lambda_2$ and $\lambda_3$ are parameters to fix the importance of them. 

\subsection{Complexity Analysis}

Let $N^t$, $b$ and $f^t$ denote the number of nodes, the batch size, and the input feature dimension of the target graph, respectively. The SEMD module estimates node-wise entropy via pairwise similarity distributions, whose computational complexity is $O(b^2f^t)$. Additionally, the computational complexity of structural constraint is $O(b^2)$. Given the number of node $N^t$, the total computational complexity of SEMD is $O\left( Nbf^t \right)$. For the domain-aware semi-hard negative sampling strategy, only $T$ informative negatives are retained for each of the \(N\) anchors. This reduces the contrastive learning complexity from the naïve \(O(N^{2}d)\) to $O(NTd)$. Overall, the computational complexity of SUCRe is $O(Nbf^t + NTf^t)$. By contrast, the overall computational complexity of the strongest baseline GraphLoRA is $O(KNbf^t + N^2f^t)$, where $K>1$ is the number of kernels and $T<N$. These demonstrate that SUCRe tends to be more efficient than GraphLoRA. Detailed computational complexity analysis of SEMD and its competitors is included in Appendix B.

\begin{table*}[!t]\centering
\renewcommand{\arraystretch}{0.98}
\caption{The accuracy of node classification of public, few-shot and different source-target transfer setting (PM, CS, C, P and Com correspond to pre-training on PubMed, CiteSeer, Cora, Photo and Computers, respectively). \label{table1}}
\begin{tabular}{clcccccccccc}
\toprule
\multirow{2}{*}{Method}
& \multirow{2}{*}{\makecell{Pre-\\Train}}
& \multicolumn{2}{c}{PubMed}
& \multicolumn{2}{c}{CiteSeer}
& \multicolumn{2}{c}{Cora}
& \multicolumn{2}{c}{Photo}
& \multicolumn{2}{c}{Computers} \\
\cmidrule(lr){3-4} \cmidrule(lr){5-6} \cmidrule(lr){7-8} \cmidrule(lr){9-10} \cmidrule(lr){11-12}

& 
&
public & 20-shot
& public & 20-shot
& public & 20-shot
& public & 20-shot
& public & 20-shot \\
\midrule

GCN
&/
& 0.7870 & 0.7206 
& 0.7120 & 0.6501  
& 0.8150 & 0.7350 
& 0.9129 & 0.8621  
& 0.8560 & 0.7504  \\ 

{PPNP}
&/
& 0.7980 & 0.7806 
& 0.7303 & 0.6815  
& 0.8195 & 0.7710 
& 0.9192 & 0.8601   
& 0.8460 & 0.7952  \\ 

{GPPT}
&/
& 0.7973 & 0.7482 
& 0.6989 & 0.6514  
& 0.8026 & 0.7518 
& 0.9210 & 0.8787   
& 0.8850 & 0.7961  \\ 

{DDSM}
&/
& 0.7905 & 0.7495 
& 0.6780 & 0.6426  
& 0.7999 & 0.7710 
& 0.9068 & 0.8785   
& 0.8751 & 0.8080  \\ 

\midrule


\multirow{5}{*}{\rotatebox[origin=c]{0}{\footnotesize{\makecell{AdapterGNN\\(\textit{AAAI, 2024})}}}}
&PM
& 0.7640 & 0.7220 
& 0.6288 & 0.5899  
& 0.7561 & 0.6519 
& 0.9037 & 0.8867   
& 0.8886 & 0.7650  \\

&CS
& 0.7411 & 0.7015 
& 0.6928 & 0.6892 
& 0.7961 & 0.7539 
& 0.9100 & 0.8906   
& 0.8603 & 0.7549  \\

&C
& 0.7381 & 0.6950 
& 0.6508  & 0.6259 
& 0.8008  & 0.7509 
& 0.9158  &  0.8891  
& 0.8691  & 0.7514  \\

&P
& 0.7269 & 0.6562 
&  0.6504  & 0.6007 
&  0.7751  & 0.6608 
& 0.9227  & 0.8933   
&  0.8752  & 0.7370  \\

&Com
& 0.7346 & 0.6089
&  0.6408  & 0.6198 
& 0.7526   & 0.6052 
& 0.9237  &  0.8998  
&  0.8802  & 0.7156  \\ 
\midrule


\multirow{5}{*}{\rotatebox[origin=c]{0}{\footnotesize{\makecell{GraphControl\\(\textit{WWW, 2024})}}}}

&PM
& 0.6934 & 0.6158 
& 0.6558 & 0.6019  
& 0.7367 & 0.6041
& 0.8919 & 0.8279   
& 0.8266 & 0.7594  \\

&CS
& 0.6613 & 0.6629
& 0.6819  & 0.5616
&  0.7362  & 0.6420
& 0.8857  & 0.8272   
&  0.8284  & 0.7471  \\

&C
& 0.7081 & 0.6283
&  0.6317  & 0.5717
&  0.7455  & 0.6752
& 0.8859  &  0.8271  
&  0.8244  & 0.7294  \\

&P& 0.6851 & 0.6111 
&  0.6680  & 0.5728
&  0.7323  & 0.6316 
& 0.8835  & 0.8625  
&  0.8258  & 0.7614  \\

&Com
& 0.6994  & 0.6539
& 0.6217  & 0.5730 
& 0.6926  & 0.6335
& 0.8864  & 0.8483  
& 0.8260  & 0.7591  \\ 
\midrule

\multirow{5}{*}{\rotatebox[origin=c]{0}{\footnotesize{\makecell{GraphLoRA\\(\textit{KDD, 2025})}}}}
&PM
& \textbf{0.8018} & \textbf{0.7692} 
& 0.7198 & 0.6910  
& \textbf{0.8146} & 0.7882 
& \textbf{0.9289} & 0.9090   
& 0.8769 & 0.8025  \\

&CS
& \underline{\textbf{0.8042}} & \underline{\textbf{0.7756}} 
&  0.7242  & 0.6968 
&  0.8156  & 0.7962 
& \textbf{0.9293}  & 0.9048   
&  0.8798  & 0.8041  \\

&C
& \textbf{0.8028} & \textbf{0.7740} 
&  0.7192  & 0.6928 
&  0.8182  & 0.7916 
& 0.9269  &  0.9094  
&  0.8742  & 0.8000  \\

&P& 0.7598 & \textbf{0.7684} 
&  0.6986  & 0.6774 
&  0.8108  & 0.7854 
& 0.9220  & 0.8921   
&  0.8749  & 0.8045  \\

&Com
& 0.7806 & \textbf{0.7394} 
&  0.6960  & 0.6356 
& 0.8162   & \textbf{0.7940} 
& 0.9197  &  0.8916  
&  0.8752  & 0.7974  \\ 
\midrule

\multirow{5}{*}{\rotatebox[origin=c]{0}{\footnotesize
{\makecell{\textbf{SUCRe}\\ (\textit{Proposed})}}}}
& PM
& 0.7837  &  0.7188   &  \textbf{0.7293}   &   \textbf{0.7108}   &  0.8127   & \textbf{0.7942} &  0.9256  & \textbf{0.9172}  & \textbf{0.8871}   & \textbf{ 0.8195}  \\

&CS
&  0.7803 & 0.6362    &  \textbf{0.7266}   &   \textbf{0.7260}  &  \textbf{0.8195}   & \underline{\textbf{0.8012}}  &  0.9243  & \textbf{0.9120}  &  \textbf{0.8821}   & \textbf{0.8135}  \\

&C
&  0.7842 & 0.7008    &  \textbf{0.7306}   & \underline{\textbf{0.7332}}     &  \underline{\textbf{0.8216}}   & \textbf{0.7962}  & \underline{\textbf{0.9295}}  & \underline{\textbf{0.9180}} &  \underline{\textbf{0.8885}}   &  \textbf{0.8239}  \\

&P
&  \textbf{0.7829} &  0.7080   &  \underline{\textbf{0.7335}}   &  \textbf{0.7192}   &  \textbf{0.8201}   & \textbf{0.7962}  & \textbf{0.9272}   & \textbf{0.9161} &  \textbf{0.8849}   &  \underline{\textbf{0.8256}}   \\

&Com
&  \textbf{0.7839} &  0.6976   & \textbf{0.7270}   &  \textbf{0.7164}    &  \textbf{0.8179}   &  0.7854  & \textbf{0.9260}  &  \textbf{0.9176}  & \textbf{0.8868}   & \textbf{0.8244} \\
\bottomrule
\end{tabular}
\end{table*}

\section{Experiments}

In this section, we conduct extensive experiments on benchmark datasets to evaluate SUCRe’s effectiveness in GTL, in particular considering the following research questions: 

\textbf{RQ. 1:} Does SUCRe outperform baselines, in public, few-shot and different source-target pair setting? 

\textbf{RQ. 2:} Does SUCRe tend to be efficient, regarding GPU memory load and time consumption?  

\textbf{RQ. 3:} How much do proposed modules contribute to SUCRe’s effectiveness? 

\subsection{Experimental Setting}

\subsubsection{Datasets. } The well-known datasets are utilized in experiments, including PubMed, Cora, CiteSeer \cite{CiteSeer}, Amazon Photo, Amazon Computers, Physics \cite{Physics}, WikiCS \cite{WikiCS}. Details for them are presented in Appendix C. 

\subsubsection{Baselines. } 
Our baselines include non-transfer methods (GCN \cite{GCN}, PPNP \cite{ppnp}, GPPT \cite{GPPT} and DDSM \cite{DDSM}) and transfer ones (AdapterGNN \cite{AdapterGNN}, GraphControl \cite{GraphControl}, GraphLoRA \cite{GraphLoRA}). Main experiments are performed on a server equipped with an Intel(R) Xeon(R) Platinum 8470Q CPU and an NVIDIA RTX 5090 GPU. Other experimental details are included in Appendix D. 


\subsection{Effectiveness. (RQ 1)} 


The performance of SUCRe and its competitors is presented in Table \ref{table1}. For each source-target transfer setting, the method achieving the best performance is highlighted in bold, while the best result in each column is underlined. Overall, SUCRe consistently ranks among the top-performing methods, achieving either the first or second position in most cases, demonstrating its superior cross-graph transfer learning capability compared with existing baselines. Compared with the state-of-the-art GTL method GraphLoRA, SUCRe achieves improvements of up to 3.4\% and 4.18\% under the public and few-shot settings, respectively. Furthermore, the average ranking of SUCRe across all experimental settings in Table \ref{table1} is 1.32, significantly outperforming the best baseline ranking of 1.9. These results validate the effectiveness of the proposed modules.

\begin{figure}[!t]
\centering
\includegraphics[width=0.4\textwidth]{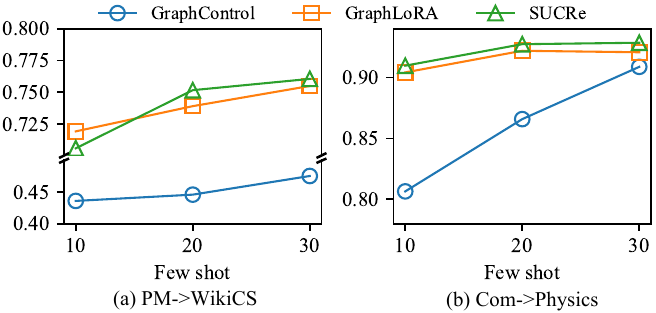} 
\caption{Results on large-scale graphs.}
\label{fig_large}
\end{figure}

In addition, we conduct experiments on the WikiCS and Physics datasets to further evaluate the scalability and effectiveness of SUCRe on large-scale graphs. As shown in Fig. \ref{fig_large}, SUCRe consistently outperforms GraphLoRA and GraphControl across different few-shot scenarios, including 10-shot, 20-shot, and 30-shot settings. These results further demonstrate the robustness and scalability of SUCRe for cross-graph transfer learning on large-scale graph data.

\begin{figure}[!t]
    \centering
    \begin{subfigure}{0.21\textwidth}
        \centering
        \includegraphics[width=\linewidth]{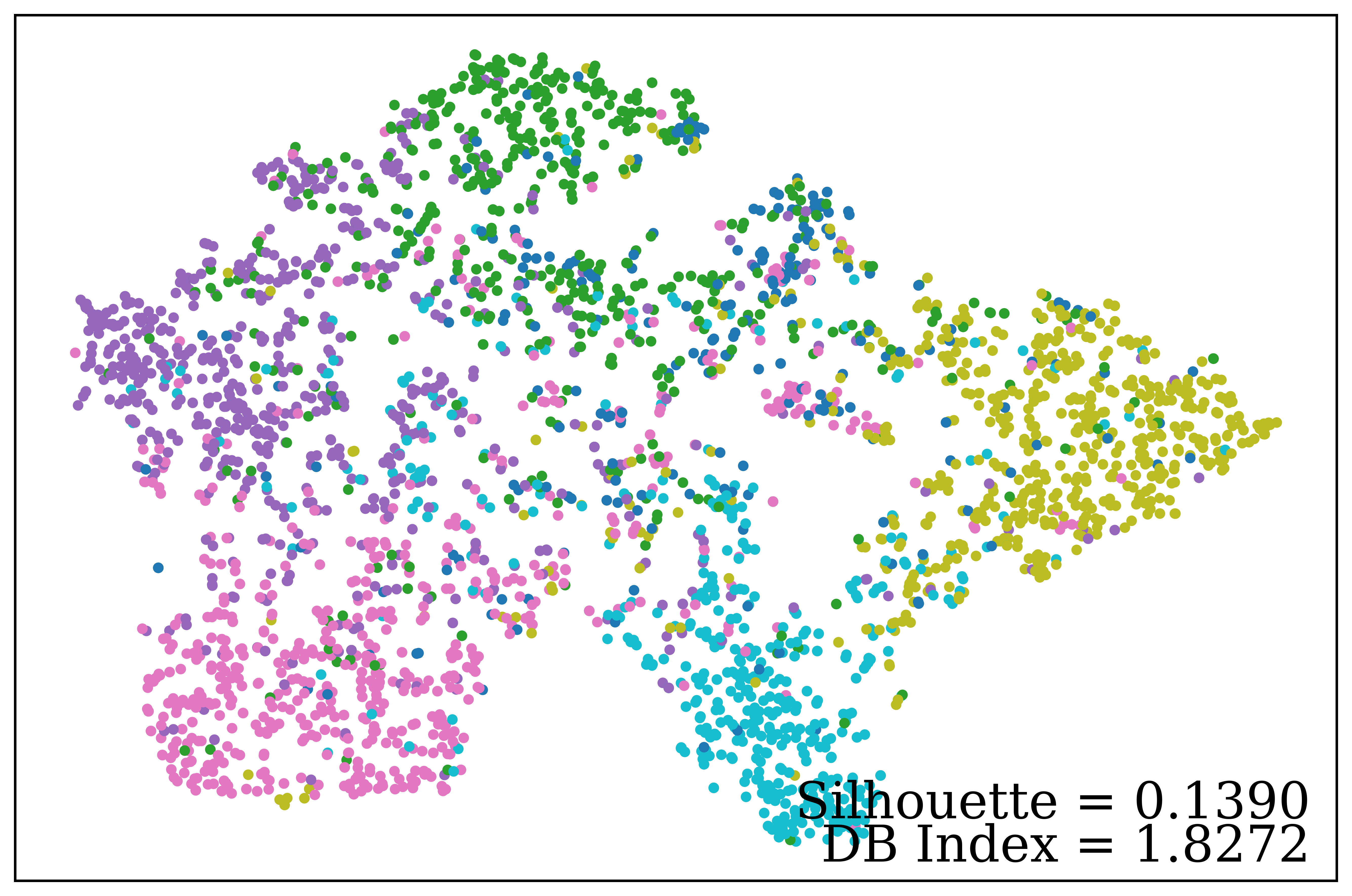}
        \caption{GCN }
        \label{fig:neg}
    \end{subfigure}
    \hfill
    \begin{subfigure}{0.21\textwidth}
        \centering
        \includegraphics[width=\linewidth]{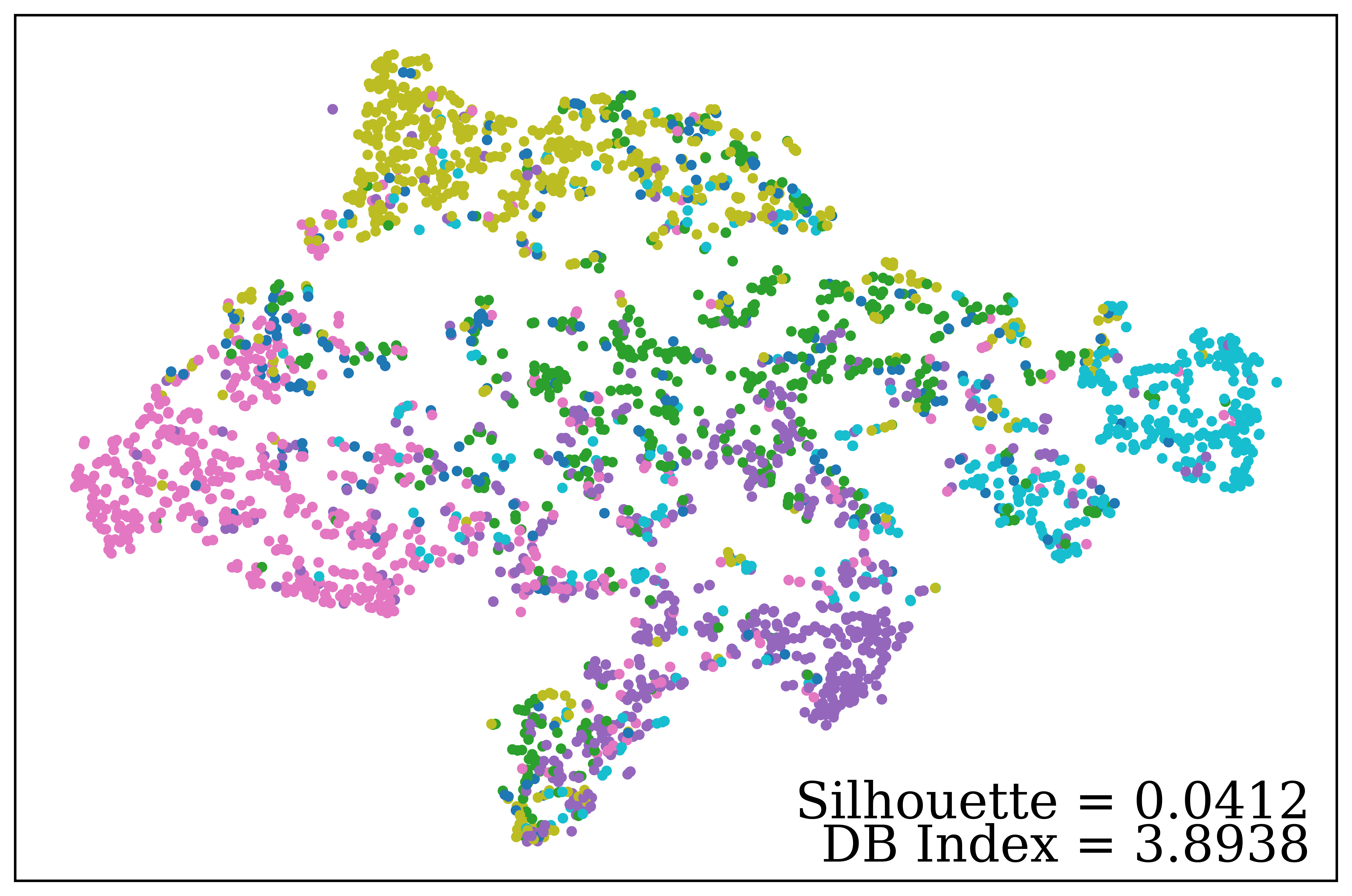}
        \caption{GraphControl}
        \label{fig:ent}
    \end{subfigure}
    \begin{subfigure}{0.21\textwidth}
        \centering
        \includegraphics[width=\linewidth]{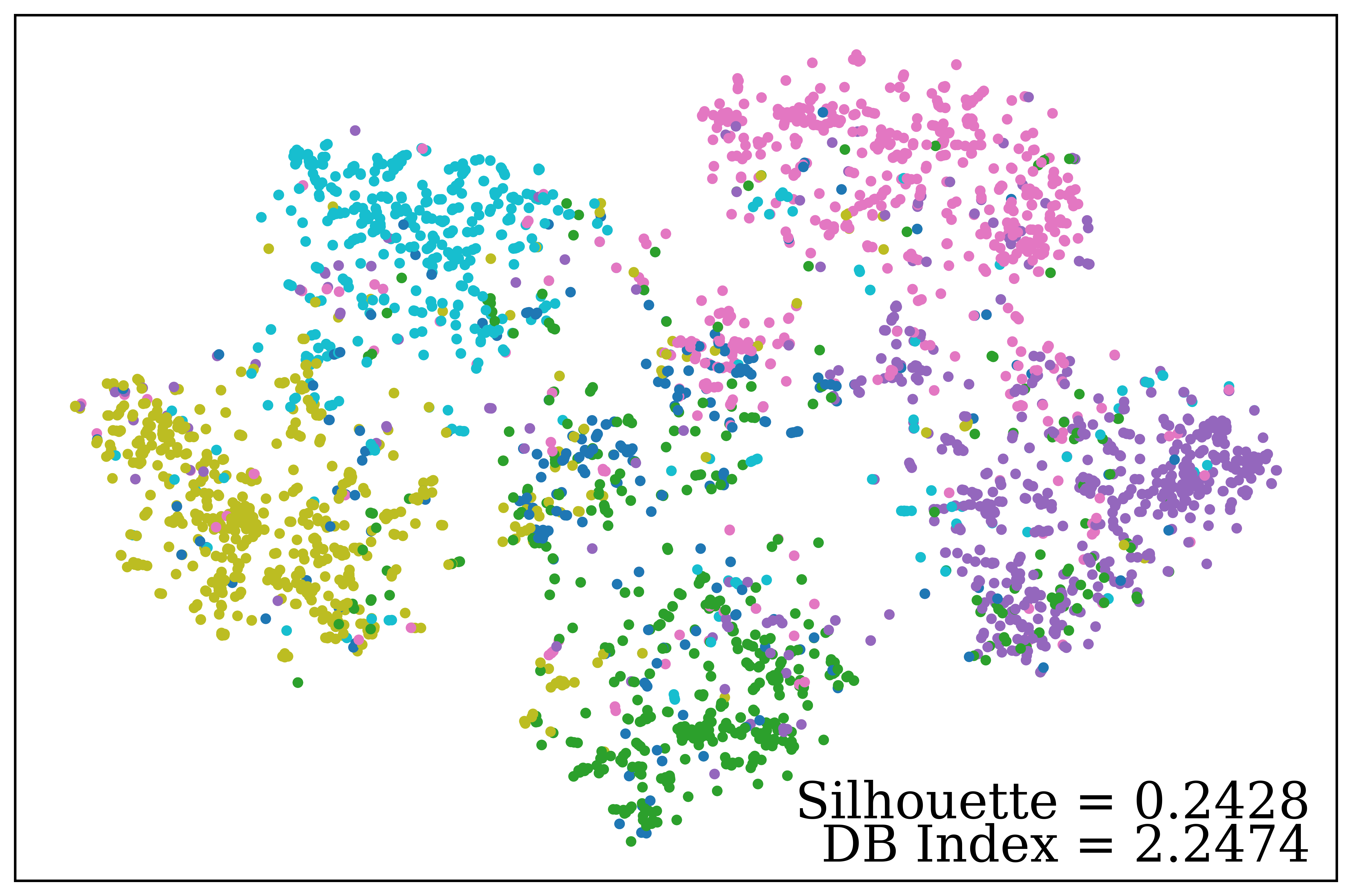}
        \caption{GraphLoRA}
        \label{fig:neg}
    \end{subfigure}
    \hfill
    \begin{subfigure}{0.21\textwidth}
        \centering
        \includegraphics[width=\linewidth]{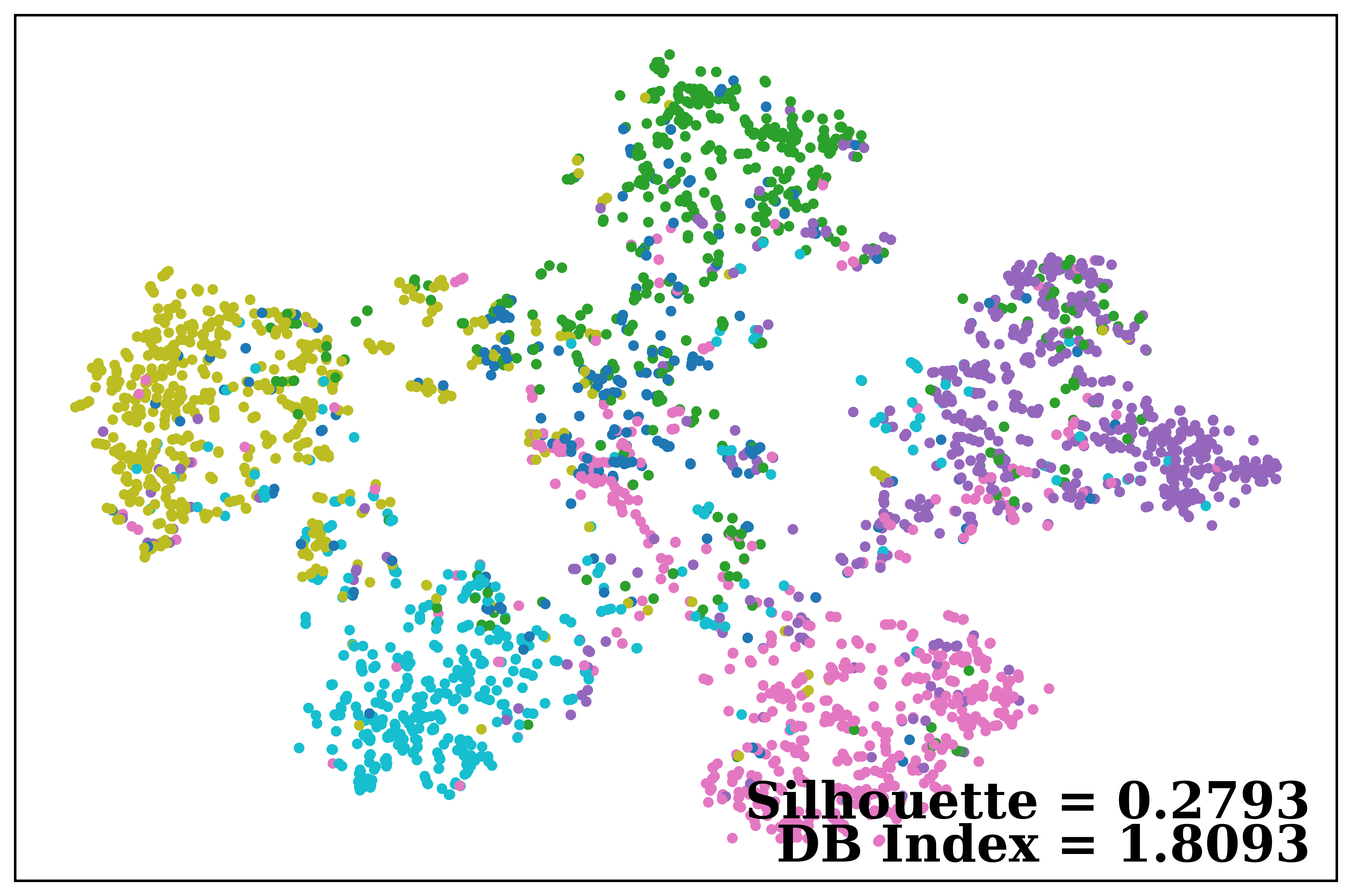}
        \caption{\textbf{SUCRe}}
        \label{fig:ent}
    \end{subfigure}
    \caption{T-SNE visualization of various methods. } 
    \label{figtsne}
\end{figure}

Furthermore, we employ the t-SNE visualization method to assess the performance of different learning methods by visualizing the learned node embeddings on CiteSeer dataset using model pre-trained on Cora dataset under public setting. As is shown in Fig. \ref{figtsne}, a node in a graph is presented as a point, with the color reflecting its label. It's easily observed that the embeddings learned by SUCRe present clearer boundaries across different classes. We also apply Silhouette score ($\uparrow$) and Davies–Bouldin (DB) index ($\downarrow$) to evaluate the performance of various methods. The $t$-SNE visualization results indicate that SUCRe achieves a Silhouette score 1.15 times higher than the strongest baseline GraphLoRA, while obtaining a DB index only 0.805 times that of the strongest baseline GraphLoRA, further validating its more discriminative representation capability via the proposed selective uncertainty-aware contrastive representation. 

\subsection{Efficiency. (RQ 2)}
\begin{figure*}[!t]
\centering
\includegraphics[width=0.76\textwidth]{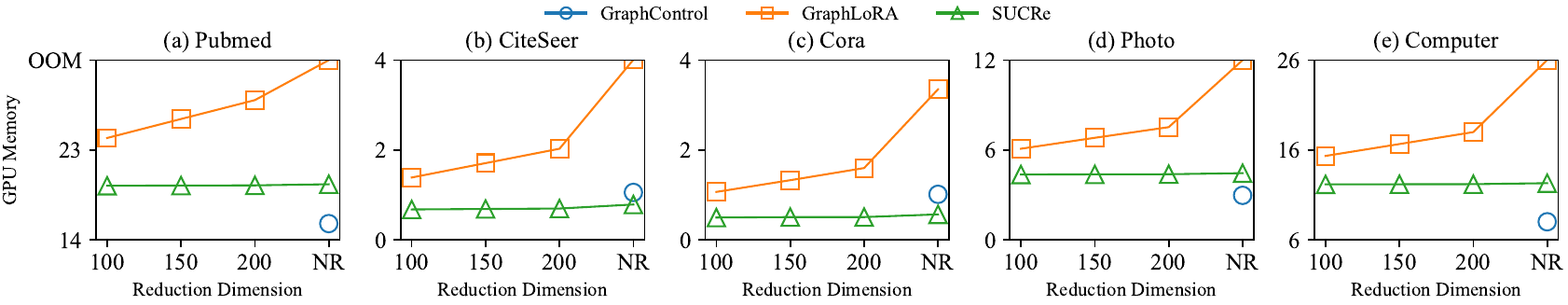} 
\caption{Memory load under various feature reduction setting using model pretrained on PubMed (`NR' means `none reduction').}
\label{fig_memory}
\end{figure*}

As discussed above, the efficiency of graph transfer learning is critically important. Therefore, we conducted experiments to evaluate the efficiency of SUCRe. The experimental results are presented in Fig. \ref{fig_memory}. First, SUCRe requires substantially less GPU memory than GraphLoRA. More importantly, as the output channel number of feature reduction increases, the memory consumption of GraphLoRA exhibits a rapid growth trend, whereas the memory usage of SUCRe remains nearly remained. These results demonstrate the significant advantage of SUCRe in terms of GPU memory load. Meanwhile, although GraphControl achieves lower memory consumption than GraphControl on large-scale graphs, its accuracy results in Table \ref{table1} reveal that such memory savings come at the cost of substantially inferior performance, making it less competitive. Overall, SUCRe not only improves prediction accuracy but also reduces the memory requirements compared with existing strong baseline methods, achieving a better trade-off between effectiveness and efficiency.

\begin{figure*}[!t]
\centering
\includegraphics[width=0.76\textwidth]{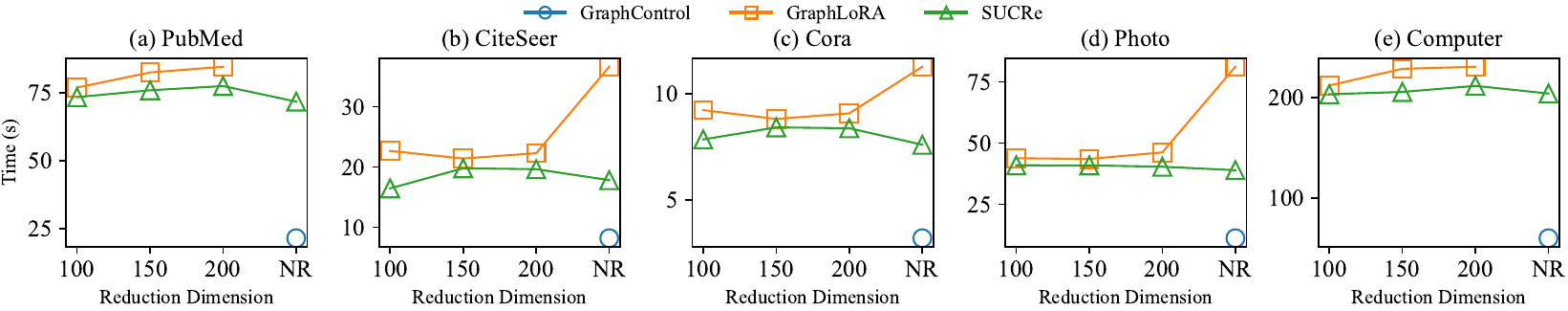} 
\caption{Time cost under various feature reduction setting using model pretrained on Cora (`NR' means `none reduction').}
\label{figtime}
\end{figure*}

\begin{table*}[!t]\centering
\renewcommand{\arraystretch}{0.98}
\caption{Ablation study with pre-trained model on $Photo$ Dataset. \label{table2}}
\begin{tabular}{lcccccccccc}
\toprule
\multirow{2}{*}{ }
& \multicolumn{2}{c}{PubMed}
& \multicolumn{2}{c}{CiteSeer}
& \multicolumn{2}{c}{Cora}
& \multicolumn{2}{c}{Photo}
& \multicolumn{2}{c}{Computers} \\
\cmidrule(lr){2-3} \cmidrule(lr){4-5} \cmidrule(lr){6-7} \cmidrule(lr){8-9} \cmidrule(lr){10-11}

& 
public & 20-shot
& public & 20-shot
& public & 20-shot
& public & 20-shot
& public & 20-shot \\
\midrule

w/o Feat. adapt. 
& 0.7620 & 0.7357 &  0.7257  & 0.6579 &  0.8031  & 0.7903 & 0.9182  & 0.9088   &  0.8715  & 0.8121  \\
w/ MMD
& 0.7695 & 0.7602 &  0.7155  & 0.6533 &  0.8085  & 0.7941 & 0.9026  & 0.9139   &  0.8840  & 0.8212  \\

w/ SMMD
& 0.7826 & \textbf{0.7804} &  0.7096  & 0.6836 &  0.7977  & 0.7935 & 0.9180  &  0.9052  &  0.8809  & 0.8228  \\

w/o Semi-hard mining
& \textbf{0.7837} & 0.6866 &  0.7331  &  0.7025 &  0.8064  & 0.7944 & 0.9201  & 0.9132   & 0.8830  & 0.8118 \\

SUCRe
& 0.7829 &  0.7080   &  \textbf{0.7335}   &  \textbf{0.7192}   &  \textbf{0.8201}   & \textbf{0.7962}  & \textbf{0.9272}   & \textbf{0.9161} &  \textbf{0.8849}   &  \textbf{0.8256} \\

\bottomrule
\end{tabular}
\end{table*}

As for time consumption, as shown in Fig. \ref{figtime}, SUCRe consistently requires less  time than GraphLoRA under the same settings, demonstrating its advantage in computational efficiency. And this phenomenon is in agreement with the complexity analysis of SUCRe. In many cases, the time consumption decrease as the reduction dimension increases. This is because the time of dimension reduction increase as the reduction dimension increases. However, this does not undermine the overall superiority of SUCRe over baselines. Although GraphControl achieves the lowest time consumption among the compared methods, its performance is accompanied by a substantial degradation in accuracy, making it less practical. Overall, although SUCRe is not always the fastest method, its overall computational cost remains highly competitive compared with existing approaches. 

\subsection{Ablation Study. (RQ 3)}

To verify each component contributes to the performance of SUCRe, we conduct ablation study, whose results are presented in Table \ref{table2}. To be specific, `w/o Feat. adapt.' means abandon the feature adaptation module completely. `w/ MMD' means replacing SEMD proposed in this work with simple MMD, while keeping semi-hard negative sample mining remained. `w/ SMMD' means replacing SEMD with simple SMMD in GraphLoRA, while keeping the other components unchanged. And `w/o semi-hard mining' means only removing the semi-hard mining module from SUCRe. As illustrated in Table \ref{table2}, it is shown that SUCRe consistently outperforms the variants in most cases, demonstrating the contribution of each module of SUCRe. 

\section{Conclusion}

GTL provides an effective solution for alleviating label scarcity by transferring knowledge across graphs. However, existing GTL methods still suffer from limited transferability beacuse of ignoring uncertainty and sample selection. In this work, we propose SUCRe, an efficient selective uncertainty-aware contrastive representation method for GTL, that incorporates uncertainty-aware feature adaptation and selective contrastive learning. The proposed uncertainty-aware feature adaptation enables adaptive feature distribution matching between source and target graphs, while the semi-hard mining module improves transfer effectiveness by selecting informative samples and reducing unnecessary optimization. Extensive experiments on benchmark datasets demonstrate that SUCRe consistently achieves competitive or superior performance compared with existing GTL methods, especially in few-shot scenarios, while maintaining lower computational and memory costs. These results highlight the effectiveness of SUCRe for scalable and efficient GTL. 

%

\end{document}